\documentclass[runningheads]{llncs}
\usepackage{graphicx}
\usepackage{xcolor}
\usepackage{bm}
\usepackage{amsmath}
\usepackage{amsfonts}
\usepackage{float}
\usepackage{booktabs}
\usepackage{multirow} 
\usepackage{url}
\usepackage{hyperref}
\usepackage{cite}
\usepackage{array}
\usepackage{verbatim}
\usepackage{color}
\usepackage{marvosym}
\usepackage[ruled,linesnumbered]{algorithm2e}
\usepackage{mathtools}

\makeatletter
\newcommand{\printfnsymbol}[1]{%
  \textsuperscript{\@fnsymbol{#1}}%
}
\makeatother
\usepackage[T1]{fontenc}
\begin{document}
%
\title{Better Generalization of White Matter Tract Segmentation to Arbitrary Datasets with Scaled Residual Bootstrap}

%

\titlerunning{Generalization of WM Tract Segmentation with Scaled Residual Bootstrap}
%
%
\author{Wan Liu\textsuperscript{[0000-0002-9641-8596]} \and Chuyang Ye\textsuperscript{[0000-0001-5839-1559]}\textsuperscript{(\Letter)}}
\authorrunning{Liu et al.}

\institute{School of Integrated Circuits and Electronics, Beijing Institute of Technology,\\ Beijing, China\\
\email{chuyang.ye@bit.edu.cn}\\}
\maketitle              
\begin{abstract}

\textit{White matter}~(WM) tract segmentation is a crucial step for brain connectivity studies.
It is performed on \textit{diffusion magnetic resonance imaging}~(dMRI), and \textit{deep neural networks}~(DNNs) have achieved promising segmentation accuracy.
Existing DNN-based methods use an annotated dataset for model training.
However, the performance of the trained model on a different test dataset may not be optimal due to distribution shift, and it is desirable to design WM tract segmentation approaches that allow better generalization of the segmentation model to arbitrary test datasets.
In this work, we propose a WM tract segmentation approach that improves the generalization with scaled residual bootstrap.
The difference between dMRI scans in training and test datasets is most noticeably caused by the different numbers of diffusion gradients and noise levels.
Since both of them lead to different \textit{signal-to-noise ratios}~(SNRs) between the training and test data, we propose to augment the training scans by adjusting the noise magnitude and develop an adapted residual bootstrap strategy for the augmentation.
First, with a dictionary-based linear representation of diffusion signals, we compute the signal residuals for the training dMRI scans, which can represent samples drawn from the noise distribution.
Then, we adapt the bootstrap procedure by scaling the residuals that are randomly drawn with replacement and adding the scaled residuals to the linear signal representation, where augmented dMRI scans with different SNRs are generated.
Finally, the augmented and original images are jointly included in model training.
Since it is difficult to know the SNR of the test data \textit{a priori}, we choose to perform the residual scaling with multiple factors.
To validate the proposed approach, two dMRI datasets were used, and the experimental results show that our method consistently improved the generalization of WM tract segmentation under various settings.

\end{abstract}

\section{Introduction}
\label{sec:intro}
\textit{White matter}~(WM) tract segmentation on \textit{diffusion magnetic resonance imaging}~(dMRI) provides a valuable quantitative tool for various brain studies~\cite{banihashemi2021opposing,girard2020cortical,toescu2021tractographic,veraart2021variability}. 
Manually delineated WM tracts are generally considered the gold standard segmentation, but the annotation process can be time-consuming and requires the expertise of experienced radiologists. 
Therefore, automated WM tract segmentation approaches are developed, which classify fiber streamlines~\cite{cook2005automated,garyfallidis2018recognition} obtained with tractography~\cite{basser2000vivo,jeurissen2019diffusion} or directly provide voxelwise labeling results~\cite{bazin2011direct,ratnarajah2014multi,ye2015segmentation}.
In particular, methods based on \textit{deep neural networks}~(DNNs) have substantially improved the accuracy of WM tract segmentation~\cite{zhang2020deep,wasserthal2018tractseg,lu2021volumetric}.
For example, Zhang et al.~\cite{zhang2020deep} group fiber streamlines into different WM tracts with a DNN that takes the spatial coordinates of the points along a fiber streamline as input; in~\cite{wasserthal2018tractseg}, fiber orientation maps extracted from dMRI scans are fed into a U-net~\cite{ronneberger2015u} to directly predict the existence of WM tracts at each voxel.

The DNN-based segmentation model is generally trained on a dataset where both dMRI scans and WM tract annotations are available.
However, the performance of the model on an arbitrary test dataset that is different from the training dataset may be degraded due to distribution shift, where the use of different numbers of diffusion gradients and different noise levels are two major contributing factors~\cite{ning2020cross}.
Since dMRI scans can be acquired with various protocols, the improvement of the generalization of WM tract segmentation models to arbitrary test data becomes an important research topic.
Although domain adaptation techniques~\cite{guan2021domain} may be applied to improve the generalization, they require access to the test data during model training, which is not guaranteed when arbitrary test data is considered, and thus they are out of scope for this work. 
To account for the different numbers of diffusion gradients between training and test datasets, in~\cite{wasserthal2018tractseg} additional training scans are obtained by subsampling the diffusion gradients of the training data, and this allows improved segmentation accuracy on test data.
However, the segmentation accuracy may still be improved by taking the \textit{signal-to-noise ratio}~(SNR) into consideration during model training.

In this work, we seek to further improve the generalization of WM tract segmentation from the perspective of SNR.\footnote{Note that the use of different numbers of diffusion gradients implicitly leads to different SNRs of measures derived from dMRI as well.} We focus on volumetric WM tract segmentation that directly obtains volumes of WM tract labels without requiring the tractography step.
We assume that by producing diverse SNRs for training data, the training data can better represent the test data, and the trained model can better generalize to the test data.
Therefore, we propose a scaled residual bootstrap strategy that augments the training scans with adjusted noise magnitude.
First, we estimate a linear dictionary-based representation of diffusion signals and compute the residuals of the representation.
These residuals are considered samples drawn from the noise distribution~\cite{jeurissen2011probabilistic}.
Then, for each diffusion gradient, the residual is drawn with replacement, and we adapt the standard residual bootstrap by scaling the residual.
The scaled residuals are added to the linear representation of diffusion signals to generate augmented dMRI scans with different SNRs.
Finally, the augmented images are used together with the original images for model training.
Since it is difficult to know the SNR of the test data \textit{a priori}, we choose to perform the residual scaling with multiple factors.
The proposed approach was evaluated on two brain dMRI datasets, where various experimental settings of training and test scans were considered. 
The results show that our method consistently improved the generalization of WM tract segmentation under these various settings.

\section{Methods}
\subsection{Problem Formulation}

Suppose we are given a set of dMRI scans from a training dataset and the set of their annotations of WM tracts.
We seek to train a WM tract segmentation model with good generalization, i.e., it performs well on an arbitrary test dataset.
Like existing volumetric WM tract segmentation approaches~\cite{wasserthal2018tractseg,liu2022volumetric}, the model input is fiber orientation maps computed from dMRI.
Two major factors that cause the difference between the training and test dMRI data are the use of different numbers of diffusion gradients and different noise levels.
Since increasing/decreasing the number of diffusion gradients also leads to increased/decreased SNRs in the fiber orientation maps, respectively, we assume that adjusting the SNR of the dMRI scans for model training can effectively improve the generalization of the trained model to other datasets.
Although existing approaches have considered SNR manipulation in the data augmentation operations of model training~\cite{wasserthal2018tractseg}, it is applied to the network input of fiber orientation maps.
As fiber orientations are orientations with unit lengths, adding realistic noise that is consistent with imaging physics to them is nontrivial.
Therefore, we seek to further explore data augmentation with SNR adjustment in model training to improve the generalization of WM tract segmentation models.



\subsection{Model Training with Scaled Residual Bootstrap}
\label{sec:residual bootstrap}
To produce training data with diverse SNRs and realistic noise distributions, we propose a scaled residual bootstrap strategy for model training.
For convenience, we denote the diffusion weighted signals at each voxel of a training dMRI scan by a vector $\bm{y}$, where $\bm{y}\in\mathbb{R}^{N_{\rm{d}}}$ and $N_{\rm{d}}$ is the number of diffusion gradients.
It has been shown that diffusion weighted signals can be linearly represented with a properly designed dictionary~\cite{merlet2012parametric,merlet2013continuous}:
\begin{eqnarray}
\bm{y}=\mathbf{D}\bm{x}+\bm{\epsilon},
\end{eqnarray}
where $\mathbf{D}\in\mathbb{R}^{{N_{\rm{d}}}\times{N_{\rm{a}}}}$ is the dictionary with $N_{\rm{a}}$ dictionary atoms, $\bm{x}\in\mathbb{R}^{N_{\rm{a}}}$ is the vector of dictionary coefficients, and $\bm{\epsilon}\in\mathbb{R}^{N_{\rm{d}}}$ represents the noise. 

If the distribution of $\bm{\epsilon}$ is known, different levels of realistic noise can be added to the noise-free linear representation to provide training data with different SNRs. 
This motivates us to adopt a residual bootstrap strategy, which provides a feasible way of approximating the noise distribution.
Then, by modifying the noise distribution, we achieve the goal of augmenting the SNR levels of training data.
There are two major steps in the proposed method, which are 1) residual computation and 2) data generation with scaled residuals.

\subsubsection{Residual Computation}
Like the standard residual bootstrap, we first estimate $\bm{x}$ with the pseudoinverse of $\mathbf{D}$:
\begin{eqnarray}
\hat{\bm{x}}=(\mathbf{D}^{\mathsf{T}}\mathbf{D})^{-1}\mathbf{D}^{\mathsf{T}}\bm{y},
\end{eqnarray}
where $\hat{\bm{x}}$ is the estimated coefficient vector. 
Then, the linear representation of the diffusion weighted signals can be estimated as 
\begin{eqnarray}
\hat{\bm{y}}=\mathbf{D}\hat{\bm{x}}=\mathbf{D}(\mathbf{D}^{\mathsf{T}}\mathbf{D})^{-1}\mathbf{D}^{\mathsf{T}}\bm{y}.
\end{eqnarray}
The residuals $\hat{\bm{\epsilon}}$ of the signal representation can be simply computed by subtracting $\hat{\bm{y}}$ from $\bm{y}$
\begin{eqnarray}
\hat{\bm{\epsilon}}=\bm{y}-\hat{\bm{y}}=(\mathbf{I}-\mathbf{D}(\mathbf{D}^{\mathsf{T}}\mathbf{D})^{-1}\mathbf{D}^{\mathsf{T}})\bm{y}.
\end{eqnarray}
Then, to ensure that the variances of the residuals $\hat{\bm{\epsilon}}$ are consistent with those of the noise $\bm{\epsilon}$, the residuals are corrected with the following normalization~\cite{davison1997bootstrap,jeurissen2011probabilistic}:
\begin{eqnarray}
\hat{\epsilon}_{i}'=\frac{\hat{\epsilon}_{i}}{\sqrt{1-h_{ii}}}.
\end{eqnarray}
Here, $\hat{\epsilon}_{i}$ is the $i$-th entry of $\hat{\bm{\epsilon}}$, $\hat{\epsilon}_{i}'$ is the corresponding corrected residual, and $h_{ii}$ is the $i$-th diagonal entry of $\mathbf{H}=\mathbf{D}(\mathbf{D}^{\mathsf{T}}\mathbf{D})^{-1}\mathbf{D}^{\mathsf{T}}$. 
The set $\mathcal{E}=\{\hat{\epsilon}_{i}'\}_{i=1}^{N_{\rm{d}}}$ of corrected residuals is then used in the bootstrap procedure that provides training data with diverse SNRs, and the procedure is described next.

\subsubsection{Data Generation with Scaled Residuals}

The corrected residuals $\mathcal{E}$ can be viewed as samples drawn from the noise distribution~\cite{davison1997bootstrap}, and in the standard residual bootstrap, they are randomly drawn with replacement and added to the linear representation $\hat{\bm{y}}$.
For our purpose of better generalization, we seek to generate samples with diverse SNRs.
Therefore, the standard bootstrap procedure is modified with a scaling operation.
Specifically, for the $i$-th diffusion gradient, we sample from $\mathcal{E}$ with replacement, and the sampled residual is denoted by $\tilde{\epsilon}_{i}$.
The vector comprising the sampled residuals for all diffusion gradients is represented as $\tilde{\bm{\epsilon}}=(\tilde{\epsilon}_{1},\ldots,\tilde{\epsilon}_{N_{\rm{d}}})$.
Then, a bootstrap signal $\tilde{\bm{y}}$ is generated as 
\begin{eqnarray}
\tilde{\bm{y}}=\hat{\bm{y}}+r\tilde{\bm{\epsilon}},
\label{eq:residual scaling}
\end{eqnarray}
where $r$ is the scaling factor that controls the magnitude of noise.
$r$ is selected from a predefined candidate set $\mathcal{R}$.
By repeating the scaled residual bootstrap in Eq.~(\ref{eq:residual scaling}) for each voxel, bootstrap diffusion weighted images can be generated.

Note that in dMRI acquisition, the $b$0 image without diffusion weighting is also acquired, and when more than one $b0$ images are available, their SNR can be adjusted as well.
We denote the $j$-th $b0$ signal at each voxel by $y_{j}^0$, and the number of $b$0 images is denoted by $N_0$.
Then, the residual $\hat{\epsilon}_{j}^{0}$ for the $j$-th $b0$ signal is calculated by
\begin{eqnarray}
\hat{\epsilon}_{j}^{0}=y_{j}^0-\bar{y}^0,
\end{eqnarray}
where $\bar{y}^0=\frac{1}{N_{0}}\sum_{j=1}^{N_{0}}y_{j}^0$ is the mean value of all $b0$ signals.
These residuals form a set $\mathcal{E}^{0}$.
For each $j$, a sample is drawn from $\mathcal{E}^{0}$ with replacement, which is denoted by $\tilde{\epsilon}_{j}^{0}$, and the bootstrap $b0$ signal is generated as
\begin{eqnarray}
\tilde{y}_{j}^0=\bar{y}^0+r\tilde{\epsilon}_{j}^0.
\label{eq:residual scaling_b0}
\end{eqnarray}
Here, $r$ has the same value as in Eq.~(\ref{eq:residual scaling}).
Eq.~(\ref{eq:residual scaling_b0}) is repeated for each voxel to obtain bootstrap $b0$ images.

After bootstrap $b0$ images and diffusion weighted images are generated, they are combined to obtain new dMRI scans with different SNRs.
These bootstrap dMRI scans are used to train the segmentation model together with the original dMRI scans based on the WM tract annotations.

\subsection{Implementation Details}

Our method is agnostic to the architecture of the segmentation model.
For demonstration, the state-of-the-art TractSeg architecture~\cite{wasserthal2018tractseg} is used as the backbone network, but other network structures~\cite{liu2022volumetric,li2020neuro4neuro} may also be applied. 
As in~\cite{wasserthal2018tractseg}, we extract fiber orientation maps from dMRI scans with \textit{constrained spherical deconvolution}~(CSD)~\cite{tournier2007robust} (for single-shell dMRI data) or \textit{multi-shell multi-tissue CSD}~(MSMT-CSD)~\cite{jeurissen2014multi} (for multi-shell dMRI data), and use these maps as network input.
At most three fiber orientations are allowed, and all WM tracts are jointly segmented~\cite{wasserthal2018tractseg}.

We use the SHORE basis\footnote{The default setting given in \url{https://dipy.org/documentation/1.4.1./reference/dipy.reconst/\#dipy.reconst.shore.ShoreModel} is used.}~\cite{merlet2013continuous} for the linear representation of diffusion signals, which is a common choice.
To generate bootstrap training data with diverse SNRs, the set of candidate scaling factors is $\mathcal{R}=\{2,3,4\}$.
Since it is difficult to predetermine the SNR of arbitrary test data, all values in $\mathcal{R}$ are used for bootstrap, and each value is used once for each training scan.

For model training, following~\cite{wasserthal2018tractseg}, we use the binary cross entropy loss function, which is minimized by Adamax~\cite{kingma2014adam} with a batch size of 56 and 300 training epochs; the initial learning rate is set to 0.001. 
We select the model that has the best segmentation accuracy on a validation dataset. Traditional data augmentation implemented online in TractSeg, such as intensity perturbation and spatial transformation, is also applied online in the proposed method.

\section{Results}
\subsection{Datasets and Experimental Settings}
We used two dMRI datasets to evaluate our method. The first one is the publicly available \textit{Human Connectome Project}~(HCP) dataset~\cite{van2013wu}, and the second one is an in-house dMRI dataset. A detailed description of the two datasets and their experimental settings is given below.

\subsubsection{The HCP Dataset}
\label{sec:data_hcp}

The dMRI scans in the HCP dataset were acquired with 270 diffusion gradients ($b$ = 1000, 2000, and 3000 s/mm$^{2}$) and an isotropic image resolution of 1.25 mm. 
18 $b0$ images were also acquired for each dMRI scan.
72 WM tracts were manually delineated for the HCP dataset\footnote{The annotations can be downloaded at \url{https://doi.org/10.5281/zenodo.1088277}.}. 
We used 100 scans in our experiments, where 55 and 15 scans were used as the training set and validation set, respectively, and the remaining 30 scans were used for testing. 
To improve the generalization of the segmentation model to different imaging protocols, in TractSeg~\cite{wasserthal2018tractseg}, subsampling of diffusion gradients was performed on the original training dMRI scans, where dMRI scans with 12 and 90 diffusion gradients associated with $b$ = 1000 s/mm$^{2}$ were generated for model training together with the original dMRI scans.\footnote{All $b0$ images were kept for these two cases.}
Here, we followed~\cite{wasserthal2018tractseg} and performed the subsampling as well for the original and bootstrap training data for model training.
For convenience, the original HCP dataset is referred to as HCP\_1.25mm\_270, and the subsampled datasets with 12 and 90 diffusion gradients are referred to as HCP\_1.25mm\_12 and HCP\_1.25mm\_90, respectively.

To evaluate the performance of the proposed method on test scans that were acquired with different protocols, we generated additional test sets from the 30 original test scans. 
First, like the training data in HCP\_1.25mm\_12 and HCP\_1.25mm\_90, the 12 and 90 diffusion gradients associated with $b$ = 1000 s/mm$^{2}$ were selected from the 30 test scans, respectively.
Second, 34 diffusion gradients associated with $b$ = 1000 s/mm$^{2}$ were selected for the test scans, so that their imaging protocol was different from the original and subsampled training data, and the images associated with this subsampling are referred to as HCP\_1.25mm\_34.
Only three $b0$ images were kept for HCP\_1.25mm\_34.
Finally, another test set HCP\_1.25mm\_36 was generated from the test scans by selecting 18 diffusion gradients associated with $b$ = 1000 s/mm$^{2}$ and 18 diffusion gradients associated with $b$ = 2000 s/mm$^{2}$, which also produced dMRI scans that used a different imaging protocol than the training data.
Only one $b0$ image was kept for HCP\_1.25mm\_36.
A summary of these different datasets is listed in Table~\ref{tab:dataset}. 

\begin{table}[!ht]
\caption{A summary of the datasets used in the experiments}
	\centering
	\resizebox{0.9\columnwidth}{!}{
		\begin{tabular}
		{>{\centering\arraybackslash}p{3cm} 
		 >{\centering\arraybackslash}p{2.5cm} 
		 >{\arraybackslash}p{3cm} 
		 >{\centering\arraybackslash}p{3cm}}
            \hline
            \hline 
			 Dataset & Resolution & Diffusion gradients& Usage\\
			\hline
		   \multirow{4}{*}{HCP$\_$1.25mm$\_$270} &\multirow{4}{*}{1.25 mm} & 90$\times$ b=1000 s/mm$^{2}$&\multirow{4}{*}{Training \& Test}\\
		    &  & 90$\times$ b=2000 s/mm$^{2}$& \\
			&  & 90$\times$ b=3000 s/mm$^{2}$&\\
			&  & 18$\times$ b=0 s/mm$^{2}$&\\
			\hline
			
		    \multirow{2}{*}{HCP$\_$1.25mm$\_$12} &\multirow{2}{*}{1.25 mm} & 12$\times$ b=1000 s/mm$^{2}$&\multirow{2}{*}{Training \& Test}\\
			&  & 18$\times$ b=0 s/mm$^{2}$&\\
			\hline
   
		    \multirow{2}{*}{HCP$\_$1.25mm$\_$90} &\multirow{2}{*}{1.25 mm} & 90$\times$ b=1000 s/mm$^{2}$&\multirow{2}{*}{Training \& Test}\\
			&  & 18$\times$ b=0 s/mm$^{2}$&\\
			\hline
   
			\multirow{2}{*}{HCP$\_$1.25mm$\_$34} &\multirow{2}{*}{1.25 mm} & 34$\times$ b=1000 s/mm$^{2}$&\multirow{2}{*}{Test}\\
		    &  & 3$\times$ b=0 s/mm$^{2}$&\\
			\hline			
			 \multirow{3}{*}{HCP$\_$1.25mm$\_$36} & \multirow{3}{*}{1.25 mm} & 18$\times$ b=1000 s/mm$^{2}$& \multirow{3}{*}{Test}\\
		    &  & 18$\times$ b=2000 s/mm$^{2}$\\
		    &  & 1$\times$ b=0 s/mm$^{2}$&\\
			\hline

			\multirow{4}{*}{IH$\_$1.7mm$\_$270} & \multirow{4}{*}{1.7 mm}  & 90$\times$ b=1000 s/mm$^{2}$&\multirow{4}{*}{Test}\\
		    &  &  90$\times$ b=2000 s/mm$^{2}$& \\
		    &  &  90$\times$ b=3000 s/mm$^{2}$&\\
		    &  & 1$\times$ b=0 s/mm$^{2}$&\\
			\hline
		    
            \multirow{3}{*}{IH$\_$1.7mm$\_$36} &  \multirow{3}{*}{1.7 mm} & 18$\times$ b=1000 s/mm$^{2} $& \multirow{3}{*}{Test}\\
		    &  &  18$\times$ b=2000 s/mm$^{2}$&\\
		    &  & 1$\times$ b=0 s/mm$^{2}$&\\
            \hline
			\hline
		\end{tabular}
	}
\label{tab:dataset}
\end{table}

In addition, to investigate the impact of the amount of training data on the segmentation, three other experimental settings were considered, where 10, 20, or 30 training subjects were used and the other settings were not changed.


\subsubsection{The In-house Dataset}
The segmentation models trained on the HCP dataset were also applied to an in-house dataset for further evaluation.
The dMRI scans in the in-house dataset were acquired with 270 diffusion gradients ($b$ = 1000, 2000, and 3000 s/mm$^{2}$) and one $b0$ image.
The spatial resolution is 1.7~mm isotropic. 
These scans were acquired on a scanner that is different from that of the HCP dataset.
Due to the annotation cost, only ten of the 72 annotated WM tracts of the HCP dataset were manually delineated, and the delineation was performed on 17 in-house dMRI scans.
These annotations were used only to evaluate the segmentation accuracy. 
This in-house dataset is referred to as IH\_1.7mm\_270.
We also synthesized another dataset IH\_1.7mm\_36 from IH\_1.7mm\_270 for evaluation, where 18 diffusion gradients of $b$ = 1000 s/mm$^{2}$ and 18 diffusion gradients of $b$ = 2000 s/mm$^{2}$ were selected from the original scans.
These two datasets are also summarized in Table~\ref{tab:dataset}.

\begin{figure}[!t]
  \centering
  \includegraphics[width=0.99\columnwidth]{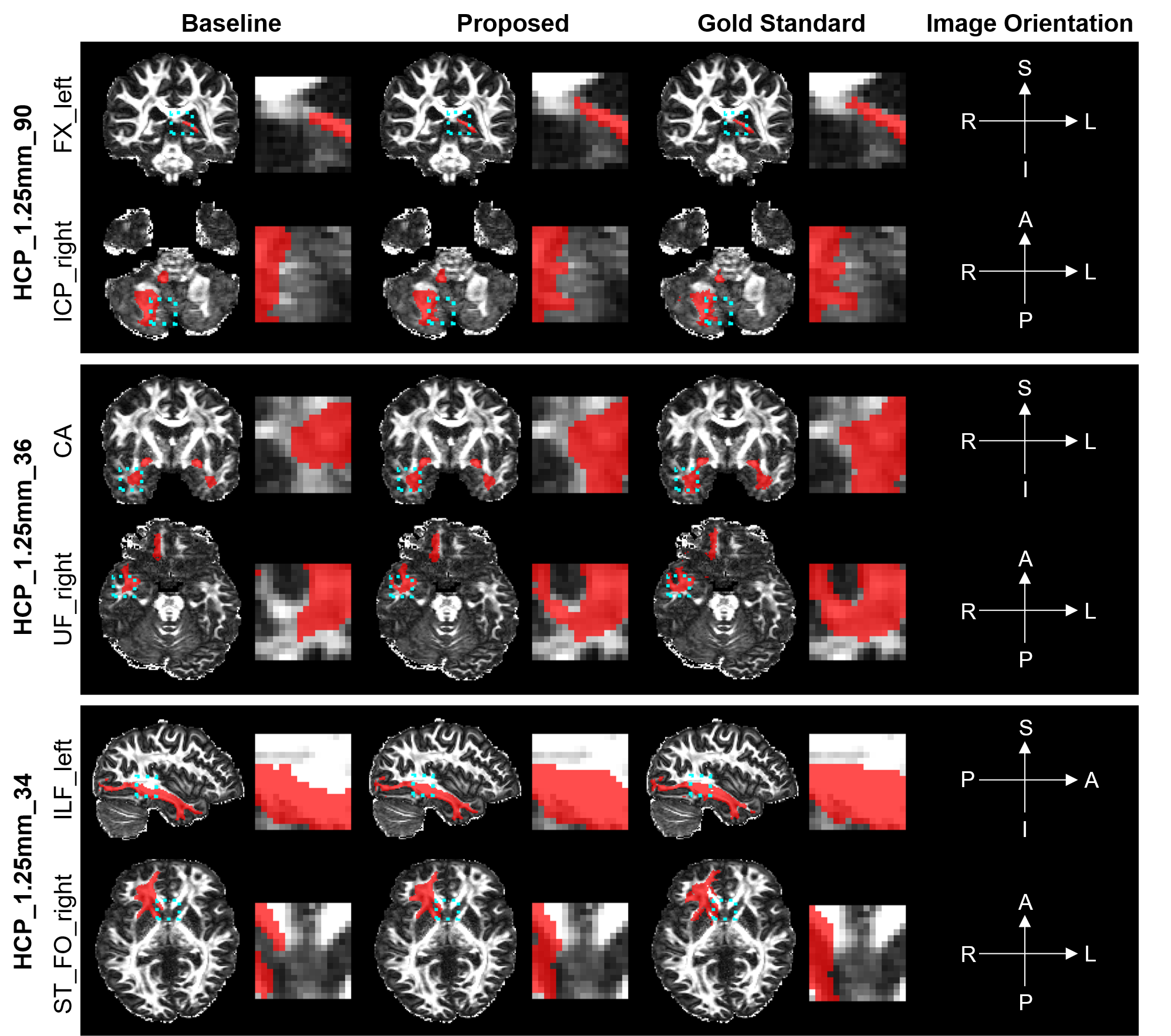}
  \caption{Representative segmentation results (red) for the HCP dataset, together with the gold standard (manual annotation) for reference. The cross-sectional views of the segmented tracts are shown, and they are overlaid on fractional anisotropy maps.
  Zoomed views of the highlighted regions are also displayed for better comparison. 
  The image orientation is shown in the rightmost column.
  For the meaning of the tract abbreviations, we refer readers to~\cite{wasserthal2018tractseg}.}
  \label{fig:qualitative comparison}
\end{figure}

\subsection{Evaluation of Segmentation Results on the HCP Dataset}

We first present the evaluation of the segmentation results on the HCP dataset.
Our method was compared with TractSeg without using bootstrap (but with the subsampling of diffusion gradients), which is referred to as the baseline method.



Examples of the segmentation results are shown in Fig.~\ref{fig:qualitative comparison}.
For demonstration, here we show the results of representative WM tracts on HCP\_1.25mm\_90, HCP\_1.25mm\_36, and HCP\_1.25mm\_34 when 55 training subjects were used. 
For reference, the gold standard (manual delineation) is also displayed. 
In Fig.~\ref{fig:qualitative comparison}, cross-sectional views of the WM tracts are given, and regions are highlighted with zoomed views for better comparison.
It can be seen that the segmented tracts of the proposed method have more similar spatial coverage to the gold standard than the baseline method.

\begin{table}[!t]
\caption{The mean Dice coefficient (\%) of all 72 WM tracts and the individual average Dice coefficients (\%) of the three most challenging tracts for the HCP dataset across different settings. The proposed method was compared with the baseline method using paired Student's $t$-tests, and asterisks indicate that the difference between the two methods is statistically significant (***: $p<0.001$).}
\centering
\resizebox{0.99\columnwidth}{!}{
\begin{tabular}
	   {>{\centering\arraybackslash}p{3cm} 
		 >{\centering\arraybackslash}p{1.5cm} 
		 >{\centering\arraybackslash}p{2cm}
		 >{\centering\arraybackslash}p{0.6cm}
		 >{\arraybackslash}p{0.8cm}
		 >{\centering\arraybackslash}p{0.6cm}
		 >{\arraybackslash}p{0.8cm}
   	 >{\centering\arraybackslash}p{0.6cm}
		 >{\arraybackslash}p{0.8cm}
		 >{\centering\arraybackslash}p{0.6cm}
		 >{\arraybackslash}p{0.8cm}}
\hline
\hline
\multirow{2}{*}{Dataset} & \multirow{2}{*}{Tract} & \multirow{2}{*}{Method} & \multicolumn{8}{c}{Number of training subjects}\\
\cline{4-11}
& & & \multicolumn{2}{c}{10} & \multicolumn{2}{c}{20} & \multicolumn{2}{c}{30}& \multicolumn{2}{c}{55}\\
\hline
\hline
\multirow{8}{*}{HCP$\_$1.25mm$\_$270} & \multirow{2}{*}{All} & Baseline&80.0 &***& 81.8&***& 82.3&*** & 83.4&***\\
& & Proposed & 80.9 & & 83.1& & 83.5 & &84.0&\\
\cline{2-11}
& \multirow{2}{*}{CA} & Baseline & 52.4&***& 61.6&***& 63.7&***& 67.3&*** \\
& & Proposed & 56.6&& 65.7&&68.0&& 69.4&\\
\cline{2-11}
& \multirow{2}{*}{FX\_left} & Baseline & 55.8&***& 67.6&***&68.1&***& 70.7&*** \\
& & Proposed & 65.0&&73.6&&73.9&& 73.7&\\
\cline{2-11}
& \multirow{2}{*}{FX\_right} & Baseline & 51.1&***& 59.5&***&61.4&***& 64.9&*** \\
& & Proposed & 55.3&& 68.0&&68.3&& 69.5&\\
\cline{2-11}
\hline
\multirow{8}{*}{HCP$\_$1.25mm$\_$90} & \multirow{2}{*}{All} & Baseline&79.0&***& 80.9&***& 81.4&*** & 82.9&***\\
& & Proposed & 80.2&& 82.8 && 83.2 && 83.7&\\
\cline{2-11}
& \multirow{2}{*}{CA} & Baseline & 51.0&***& 61.1&***&63.6&***& 66.3&*** \\
& & Proposed & 56.9&& 65.1&&67.4&& 68.7&\\
\cline{2-11}
& \multirow{2}{*}{FX\_left} & Baseline & 55.6&***& 63.5&***&65.9&***& 68.9&*** \\
& & Proposed & 62.8&& 72.2&&72.6&& 72.6&\\
\cline{2-11}
& \multirow{2}{*}{FX\_right} & Baseline & 47.8&***& 57.9&***&57.1&***& 63.6&*** \\
& & Proposed & 56.4&& 67.5&&67.1&& 68.1&\\
\cline{2-11}
\hline
\multirow{8}{*}{HCP$\_$1.25mm$\_$12} & \multirow{2}{*}{All} & Baseline&77.9&***& 80.2&***& 80.8&*** & 82.2&***\\
& & Proposed & 79.7&& 82.4&& 82.8 && 83.4&\\
\cline{2-11}
& \multirow{2}{*}{CA} & Baseline & 47.7&***& 59.2&***&61.9&***& 64.8&*** \\
& & Proposed & 56.2&& 64.1&&66.8&& 68.0&\\
\cline{2-11}
& \multirow{2}{*}{FX\_left} & Baseline & 50.6&***& 61.8&***&62.9&***& 65.5&*** \\
& & Proposed & 58.5&& 71.6&&71.9&& 72.3&\\
\cline{2-11}
& \multirow{2}{*}{FX\_right} & Baseline & 40.6&***& 54.2&***&53.3&***& 59.6&*** \\
& & Proposed & 51.4&& 66.1&&65.6&& 66.5&\\
\cline{2-11}
\hline
\multirow{8}{*}{HCP$\_$1.25mm$\_$36} & \multirow{2}{*}{All} & Baseline&79.2&***& 80.9&***& 81.5&*** &82.7&***\\
& & Proposed & 80.6&& 82.9&& 83.3 && 83.9&\\
\cline{2-11}
& \multirow{2}{*}{CA} & Baseline & 48.6&***& 60.0&***&61.6&***& 65.4&*** \\
& & Proposed & 55.2&& 65.2&&67.4&& 68.7&\\
\cline{2-11}
& \multirow{2}{*}{FX\_left} & Baseline & 53.9&***& 66.3&***&65.9&***& 68.4&*** \\
& & Proposed & 63.5&& 72.9&&73.2&& 73.5&\\
\cline{2-11}
& \multirow{2}{*}{FX\_right} & Baseline & 46.3&***& 56.7&***&58.8&***& 62.6&*** \\
& & Proposed & 52.3&& 66.9&&67.3&& 68.5&\\
\cline{2-11}
\hline
\multirow{8}{*}{HCP$\_$1.25mm$\_$34} & \multirow{2}{*}{All} & Baseline&78.4&***& 80.6&***& 81.1&*** & 82.6&***\\
& & Proposed & 80.1&& 82.7& &83.1& & 83.6&\\
\cline{2-11}
& \multirow{2}{*}{CA} & Baseline & 49.3&***& 60.3&***&63.3&***& 65.9&*** \\
& & Proposed & 56.6&&65.3&&67.3&& 68.8&\\
\cline{2-11}
& \multirow{2}{*}{FX\_left} & Baseline & 51.6&***& 61.5&***&64.3&***& 67.2&*** \\
& & Proposed & 61.1&& 72.1&&72.0&& 72.2&\\
\cline{2-11}
& \multirow{2}{*}{FX\_right} & Baseline & 43.9&***& 55.3&***&54.6&***& 62.0&*** \\
& & Proposed & 54.0&& 66.5&&66.2&& 67.5&\\
\cline{2-11}
\hline
\hline

\end{tabular}
}
\label{tab:meandice_hcp}
\end{table}

We then quantitatively evaluated the proposed method by computing the Dice coefficient between the segmentation results and the gold standard.
The mean Dice coefficient of all 72 WM tracts for each test dataset and each number of training subjects is shown in Table~\ref{tab:meandice_hcp}.
As some WM tracts can be more challenging to segment~\cite{liu2022volumetric} and the improvement of the segmentation of these tracts is important, in Table~\ref{tab:meandice_hcp} we also show the individual average Dice coefficients of the three most challenging WM tracts, which are the anterior commissure (CA), left fornix (FX\_left), and right fornix (FX\_right)~\cite{wasserthal2018tractseg,liu2022volumetric}.
Compared with the baseline method, the proposed method can consistently improve the Dice coefficients across the different cases, and the improvement is more prominent for the three most challenging WM tracts.
In addition, the Dice coefficients of the proposed method were compared with those of the baseline method using paired Student's $t$-tests, and the $p$-values are listed in Table~\ref{tab:meandice_hcp}.
It can be seen that the improvement of the proposed method is statistically significant in all cases.

By comparing the results achieved with different numbers of training subjects, we observe that the overall improvement of the proposed method tends to be greater when the number is moderate (20 and 30) than when the number is small (10) or large (55).
Moreover, the Dice coefficients of the proposed method obtained with 20 training subjects are comparable to or higher than the baseline performance achieved with 55 training subjects.
Also, when the number of training subjects increases from 20 to 30 or 55, the Dice coefficients of the proposed method are relatively stable, whereas the Dice coefficients of the baseline method can still increase.
This is possibly because the proposed method augments the training data and thus reduces the requirement for manual annotation.

\begin{table}[!t]
\caption{The mean Dice coefficient (\%) of all ten annotated WM tracts and the individual average Dice coefficients (\%) of two challenging tracts for the in-house dataset across different settings. The proposed method was compared with the baseline method using paired Student's $t$-tests, and asterisks indicate that the difference between the two methods is statistically significant (***: $p<0.001$, **: $p<0.01$, *: $p<0.05$, n.s.: $p\geq0.05$).}
\centering
\resizebox{0.99\columnwidth}{!}{
\begin{tabular}
	   {>{\centering\arraybackslash}p{3cm} 
		 >{\centering\arraybackslash}p{1.5cm} 
		 >{\centering\arraybackslash}p{2cm}
		 >{\centering\arraybackslash}p{0.6cm}
		 >{\arraybackslash}p{0.8cm}
		 >{\centering\arraybackslash}p{0.6cm}
		 >{\arraybackslash}p{0.8cm}
   	 >{\centering\arraybackslash}p{0.6cm}
		 >{\arraybackslash}p{0.8cm}
		 >{\centering\arraybackslash}p{0.6cm}
		 >{\arraybackslash}p{0.8cm}}
\hline
\hline
\multirow{2}{*}{Dataset} & \multirow{2}{*}{Tract} & \multirow{2}{*}{Method} & \multicolumn{8}{c}{Number of training subjects}\\
\cline{4-11}
& & & \multicolumn{2}{c}{10} & \multicolumn{2}{c}{20} & \multicolumn{2}{c}{30}& \multicolumn{2}{c}{55}\\
\hline
\hline
\multirow{6}{*}{IH$\_$1.7mm$\_$270} & \multirow{2}{*}{All} & Baseline&58.7&n.s.& 60.4&**&61.2&** & 61.7&n.s.\\
& & Proposed & 59.0&& 62.0&& 61.9 && 61.9&\\
\cline{2-11}
& \multirow{2}{*}{UF\_left} & Baseline & 48.4&*** & 50.7&*** & 53.1&** & 55.1 &n.s.\\
& & Proposed & 51.9 &&59.3 &&56.7&& 57.1& \\
\cline{2-11}
& \multirow{2}{*}{UF\_right} & Baseline & 52.9&n.s. & 56.5&*** & 57.3&*** & 59.1 &***\\
& & Proposed & 53.4 &&61.3 &&60.1&& 61.4& \\
\hline

\multirow{6}{*}{IH$\_$1.7mm$\_$36} & \multirow{2}{*}{All} & Baseline&57.2&**& 58.7&***& 59.3&*** & 60.4&**\\
& & Proposed& 57.8&& 61.5&& 61.5 &&61.3&\\
\cline{2-11}
\cline{2-11}
& \multirow{2}{*}{UF\_left} & Baseline & 46.2&n.s. & 46.3&*** & 47.3&*** & 51.6 &**\\
& & Proposed & 47.9 &&58.0 &&55.7&& 55.3& \\
\cline{2-11}
& \multirow{2}{*}{UF\_right} & Baseline & 48.9&*** & 54.2&*** & 53.6&*** & 56.3 &***\\
& & Proposed & 51.2 &&59.1 &&58.5&& 60.5& \\
\hline
\hline
\end{tabular}
}
\label{tab:meandice_bt}
\end{table}


\subsection{Evaluation of Segmentation Results on the In-house Dataset}

The proposed method was next applied to the in-house test datasets IH\_1.7mm\_270 and IH\_1.7mm\_36, and the mean Dice coefficients of all ten annotated WM tracts are summarized in Table~\ref{tab:meandice_bt}.
In addition, the individual average Dice coefficients of two challenging tracts, the left uncinate fasciculus (UF\_left) and right uncinate fasciculus (UF\_right)~\cite{liu2022volumetric}, are also shown in Table~\ref{tab:meandice_bt}.\footnote{CA, FX\_left, and FX\_right were not annotated for the in-house dataset.}
In each case, the proposed method achieves a higher Dice coefficient than the baseline method, and the improvement is more prominent for the two challenging tracts and for IH\_1.7mm\_36 that has a smaller number of diffusion gradients.
We also performed paired Student's $t$-tests to compare the two methods in Table~\ref{tab:meandice_bt}, and the difference between the proposed and competing methods is statistically significant in most cases.

Like the results on the HCP dataset, the improvement of the proposed method over the baseline method is greater when the number of training subjects is 20 or 30 than 10 or 55, and its performance becomes stable after the number of training subjects reaches 20.
Also, the Dice coefficients of the proposed method obtained with 20 training subjects are already better than the baseline performance achieved with 55 training subjects.



\section{Conclusion}

We have proposed a WM tract segmentation approach that better generalizes to arbitrary test datasets. In the proposed method a scaled residual bootstrap strategy is developed, where the SNR levels of the training data are adjusted based on the residuals of a linear signal representation.
This reduces the discrepancy between training and test data and thus improves the generalization of the trained segmentation model.
Our method was validated on public and in-house datasets under various data settings, and the results show that it consistently improved the segmentation accuracy in the different cases.

\subsubsection{Acknowledgements}
This work is supported by the Fundamental Research Funds for the Central Universities.

\bibliographystyle{splncs04}
\bibliography{ref}

\begin{thebibliography}{10}
\providecommand{\url}[1]{\texttt{#1}}
\providecommand{\urlprefix}{URL }
\providecommand{\doi}[1]{https://doi.org/#1}

\bibitem{banihashemi2021opposing}
Banihashemi, L., Peng, C.W., Verstynen, T., Wallace, M.L., Lamont, D.N.,
  Alkhars, H.M., Yeh, F.C., Beeney, J.E., Aizenstein, H.J., Germain, A.:
  Opposing relationships of childhood threat and deprivation with stria
  terminalis white matter. Human Brain Mapping  \textbf{42}(8),  2445--2460
  (2021)

\bibitem{basser2000vivo}
Basser, P.J., Pajevic, S., Pierpaoli, C., Duda, J., Aldroubi, A.: In vivo fiber
  tractography using {DT-MRI} data. Magnetic Resonance in Medicine
  \textbf{44}(4),  625--632 (2000)

\bibitem{bazin2011direct}
Bazin, P.L., Ye, C., Bogovic, J.A., Shiee, N., Reich, D.S., Prince, J.L., Pham,
  D.L.: {Direct segmentation of the major white matter tracts in diffusion
  tensor images}. NeuroImage  \textbf{58}(2),  458--468 (2011)

\bibitem{cook2005automated}
Cook, P.A., Zhang, H., Avants, B.B., Yushkevich, P., Alexander, D.C., Gee,
  J.C., Ciccarelli, O., Thompson, A.J.: An automated approach to
  connectivity-based partitioning of brain structures. In: International
  Conference on Medical Image Computing and Computer-Assisted Intervention. pp.
  164--171 (2005)

\bibitem{davison1997bootstrap}
Davison, A.C., Hinkley, D.V.: Bootstrap methods and their application. No.~1,
  Cambridge University Press (1997)

\bibitem{garyfallidis2018recognition}
Garyfallidis, E., C{\^o}t{\'e}, M.A., Rheault, F., Sidhu, J., Hau, J., Petit,
  L., Fortin, D., Cunanne, S., Descoteaux, M.: {Recognition of white matter
  bundles using local and global streamline-based registration and clustering}.
  NeuroImage  \textbf{170},  283--295 (2018)

\bibitem{girard2020cortical}
Girard, G., Caminiti, R., Battaglia-Mayer, A., St-Onge, E., Ambrosen, K.S.,
  Eskildsen, S.F., Krug, K., Dyrby, T.B., Descoteaux, M., Thiran, J.P.,
  Innocenti, G.M.: {On the cortical connectivity in the macaque brain: A
  comparison of diffusion tractography and histological tracing data}.
  NeuroImage  \textbf{221},  117201 (2020)

\bibitem{guan2021domain}
Guan, H., Liu, M.: Domain adaptation for medical image analysis: A survey. IEEE
  Transactions on Biomedical Engineering  \textbf{69}(3),  1173--1185 (2021)

\bibitem{jeurissen2019diffusion}
Jeurissen, B., Descoteaux, M., Mori, S., Leemans, A.: {Diffusion MRI fiber
  tractography of the brain}. NMR in Biomedicine  \textbf{32}(4),  e3785 (2019)

\bibitem{jeurissen2011probabilistic}
Jeurissen, B., Leemans, A., Jones, D.K., Tournier, J.D., Sijbers, J.:
  Probabilistic fiber tracking using the residual bootstrap with constrained
  spherical deconvolution. Human Brain Mapping  \textbf{32}(3),  461--479
  (2011)

\bibitem{jeurissen2014multi}
Jeurissen, B., Tournier, J.D., Dhollander, T., Connelly, A., Sijbers, J.:
  {Multi-tissue constrained spherical deconvolution for improved analysis of
  multi-shell diffusion MRI data}. NeuroImage  \textbf{103},  411--426 (2014)

\bibitem{kingma2014adam}
Kingma, D.P., Ba, J.: {Adam: A method for stochastic optimization}. arXiv
  preprint arXiv:1412.6980  (2014)

\bibitem{li2020neuro4neuro}
Li, B., de~Groot, M., Steketee, R.M., Meijboom, R., Smits, M., Vernooij, M.W.,
  Ikram, M.A., Liu, J., Niessen, W.J., Bron, E.E.: {Neuro4Neuro: A neural
  network approach for neural tract segmentation using large-scale
  population-based diffusion imaging}. NeuroImage  \textbf{218},  116993 (2020)

\bibitem{liu2022volumetric}
Liu, W., Lu, Q., Zhuo, Z., Li, Y., Duan, Y., Yu, P., Qu, L., Ye, C., Liu, Y.:
  Volumetric segmentation of white matter tracts with label embedding.
  NeuroImage  \textbf{250},  118934 (2022)

\bibitem{lu2021volumetric}
Lu, Q., Li, Y., Ye, C.: Volumetric white matter tract segmentation with nested
  self-supervised learning using sequential pretext tasks. Medical Image
  Analysis  \textbf{72},  102094 (2021)

\bibitem{merlet2012parametric}
Merlet, S., Caruyer, E., Deriche, R.: {Parametric dictionary learning for
  modeling EAP and ODF in diffusion MRI}. In: International Conference on
  Medical Image Computing and Computer-Assisted Intervention. pp. 10--17.
  Springer (2012)

\bibitem{merlet2013continuous}
Merlet, S.L., Deriche, R.: {Continuous diffusion signal, EAP and ODF estimation
  via compressive sensing in diffusion MRI}. Medical Image Analysis
  \textbf{17}(5),  556--572 (2013)

\bibitem{ning2020cross}
Ning, L., Bonet-Carne, E., Grussu, F., Sepehrband, F., Kaden, E., Veraart, J.,
  Blumberg, S.B., Khoo, C.S., Palombo, M., Kokkinos, I., et~al.: Cross-scanner
  and cross-protocol multi-shell diffusion {MRI} data harmonization: Algorithms
  and results. NeuroImage  \textbf{221},  117128 (2020)

\bibitem{ratnarajah2014multi}
Ratnarajah, N., Qiu, A.: {Multi-label segmentation of white matter structures:
  Application to neonatal brains}. NeuroImage  \textbf{102},  913--922 (2014)

\bibitem{ronneberger2015u}
Ronneberger, O., Fischer, P., Brox, T.: {U-net: Convolutional networks for
  biomedical image segmentation}. In: International Conference on Medical Image
  Computing and Computer-Assisted Intervention. pp. 234--241. Springer (2015)

\bibitem{toescu2021tractographic}
Toescu, S.M., Hales, P.W., Kaden, E., Lacerda, L.M., Aquilina, K., Clark, C.A.:
  Tractographic and microstructural analysis of the
  dentato-rubro-thalamo-cortical tracts in children using diffusion {MRI}.
  Cerebral Cortex  \textbf{31}(5),  2595--2609 (2021)

\bibitem{tournier2007robust}
Tournier, J.D., Calamante, F., Connelly, A.: Robust determination of the fibre
  orientation distribution in diffusion {MRI}: Non-negativity constrained
  super-resolved spherical deconvolution. NeuroImage  \textbf{35}(4),
  1459--1472 (2007)

\bibitem{van2013wu}
Van~Essen, D.C., Smith, S.M., Barch, D.M., Behrens, T.E., Yacoub, E., Ugurbil,
  K., {Wu-Minn HCP Consortium}: The {WU-Minn} human connectome project: An
  overview. NeuroImage  \textbf{80},  62--79 (2013)

\bibitem{veraart2021variability}
Veraart, J., Raven, E.P., Edwards, L.J., Weiskopf, N., Jones, D.K.: {The
  variability of MR axon radii estimates in the human white matter}. Human
  Brain Mapping  \textbf{42}(7),  2201--2213 (2021)

\bibitem{wasserthal2018tractseg}
Wasserthal, J., Neher, P., Maier-Hein, K.H.: {TractSeg - Fast and accurate
  white matter tract segmentation}. NeuroImage  \textbf{183},  239--253 (2018)

\bibitem{ye2015segmentation}
Ye, C., Yang, Z., Ying, S.H., Prince, J.L.: {Segmentation of the cerebellar
  peduncles using a random forest classifier and a multi-object geometric
  deformable model: Application to spinocerebellar ataxia type 6}.
  Neuroinformatics  \textbf{13}(3),  367--381 (2015)

\bibitem{zhang2020deep}
Zhang, F., Karayumak, S.C., Hoffmann, N., Rathi, Y., Golby, A.J., O'Donnell,
  L.J.: {Deep white matter analysis (DeepWMA): Fast and consistent tractography
  segmentation}. Medical Image Analysis  \textbf{65},  101761 (2020)

\end{thebibliography}
\end{document}